\documentclass[12pt]{article}

\newcommand\ddfrac[2]{\frac{\displaystyle #1}{\displaystyle #2}}

\usepackage{dirtytalk}
\usepackage{graphicx}
\graphicspath{ {images/} }
\usepackage{float}
\usepackage{times}
\usepackage{lineno}
\usepackage{xcolor}
\usepackage{amsmath}
\usepackage{amsfonts} 
\usepackage{soul}
\usepackage[normalem]{ulem}

\usepackage{hyperref}

\newenvironment{sciabstract}{%
\begin{quote} \bf}
{\end{quote}}

\title{A comprehensive generalization\\of the Friendship Paradox\\to weights and attributes}

\author
{Anna Evtushenko,$^{1\ast}$ Jon Kleinberg$^{1, 2}$\\
\\
\normalsize{$^{1}$Department of Information Science, Cornell University, Ithaca NY, USA}\\
\normalsize{$^{2}$Department of Computer Science, Cornell University, Ithaca NY, USA}\\
\\
\normalsize{$^\ast$anna@infosci.cornell.edu}
}

\date{}

\begin{document}

\baselineskip24pt


\maketitle
\begin{sciabstract}
The Friendship Paradox is a simple and powerful statement about node
degrees in a graph \cite{feld1991your}. However, it only applies to
undirected graphs with no edge weights, and the only node
characteristic it concerns is degree. Since many social networks are
more complex than that, it is useful to generalize this phenomenon, if
possible, and a number of papers have proposed 
different generalizations. 
Here, we unify these generalizations in a common framework, 
retaining the focus on undirected graphs and allowing for weighted edges and for numeric
node attributes other than degree to be considered, since this
extension allows for a clean characterization and links to the
original concepts most naturally. 
While the original Friendship
Paradox and the Weighted Friendship Paradox hold for all graphs,
considering non-degree attributes actually makes the extensions fail
around 50\% of the time, given random attribute assignment. We provide
simple correlation-based rules to see whether an attribute-based
version of the paradox holds. In addition to theory, our simulation
and data results show how all the concepts can be applied to synthetic
and real networks. Where applicable, we draw connections to prior work
to make this an accessible and comprehensive paper that lets one
understand the math behind the Friendship Paradox and its basic
extensions.
\end{sciabstract}

\section{Introduction}
The Friendship Paradox is a non-plural term, so it may initially be surprising that it actually comes in two versions: where friends' degrees are aggregated at the network level, naturally called \say{network-level,} and where friends' degrees are aggregated at the node level, naturally called \say{node-level.} To introduce the phenomenon broadly, the Friendship Paradox claims that in a simple connected undirected finite graph, the mean second-order degree, calculated at either the network or the node level, is no less than the first-order degree, with equality achieved only if the graph is regular.

It may be easy to get confused by the exact mechanics of the two phenomena from that broad introduction alone, so as part of this paper we formalize the concepts with algebra and make things straightforward to compute; this also makes introducing extensions easier. We call the network-level version the \say{list} version, and the node-level version the \say{singular} version---both names tie directly to how things are computed. To make it easy to see whether a phenomenon holds or fails, we look at the \say{gap} between a mean second-order quantity and a mean first-order quantity as the single computation result. A version of the Friendship Paradox (including extensions) fails if its respective gap is negative and holds otherwise. The size of the gap doesn't affect that, so we are mostly concerned with the gap sign.

The original Friendship Paradox (which we also refer
to as the FP) looks at node degrees and friend
degrees. It is most often considered for undirected graphs, which is our focus too. (Directed versions of FP have been studied as well \cite{hodas2013friendship,berenhaut2019friendship}, but they are much less tractable; to make our paper comprehensive, we explore this in the Supplementary Information.)

The first extension of the basic phenomenon allows one to
look beyond degrees and instead consider arbitrary numeric node
attributes. It is natural, for instance, to ask whether the fact that
friends have larger degrees, on average, may also somehow imply or be
correlated with them potentially having more likes on their photos,
with \say{total number of likes on one's photos} being an example of a numeric
node attribute. In the literature, this extension is called the
Generalized Friendship Paradox (GFP)~\cite{eom2014generalized}, and its relevance to comparisons of non-degree attributes
between people in social contexts has been considered by several papers 
\cite{ugander2011anatomy,momeni2016qualities}.
To make the generalization more explicit, here we can refer to this
concept as the Attribute-based Friendship Paradox (AFP). It also comes
in a list version and in a singular version, which we refer
to as the LAFP and SAFP, and which have been studied by
\cite{eom2014generalized} and 
\cite{cantwell2021friendship,evtushenko2023node} respectively.
We use the word {\em attribute} to denote the node characteristic
we are looking at in all cases---in the case of AFP, the attribute is
arbitrary; in the case of FP, the attribute is equal to node degree.

We can also suggest that each edge weight in a graph doesn't have to
be equal to $1$ (like it does in a simple graph), but is instead a
positive integer.
The distinction between the list and singular 
forms of the Friendship Paradox naturally
carries over to the weighted case, and so in our framework we
define both the LWFP and the SWFP.
In this case, edge weights would
naturally be taken into account, and the attributes would be equal to
the weighted degrees. Edge weights would also be used in second-order
calculations. The singular version of WFP has been studied in \cite{berenhaut2019friendship}.

Finally, if we allow for edges to be weighted \textbf{and} for
attributes to be unequal to degrees, we get two versions of the
Weighted Attribute-based Friendship Paradox, a formulation
that brings together aspects of these different generalizations. 
The attributes would be
arbitrary, and edge weights would be considered in second-order
calculations.

First, we introduce and discuss the two original versions of the Friendship Paradox, LFP and SFP. Then, we allow for edges to be weighted and discuss LWFP and SWFP. Finally, we allow attributes to be arbitrary on top of that and discuss LWAFP and SWAFP. In the interest of space, we discuss the unweighted attribute-based versions LAFP and SAFP in the Supplementary Information; LWAFP and SWAFP work perfectly well for the cases of no edge weights if we just take all edge weights to be 1. Because LWAFP and SWAFP consider an extension to both edge weights and attributes at once, we also call these two versions the Extended Friendship Paradox, or LEFP and SEFP. As mentioned, these work for the case of no edge weights; we will also discuss how to apply them to the case of no arbitrary node attributes.

To reiterate, the two original forms of the paradox 
combined with two independent sources of generalization---attributes and edge weights---mean that we have eight versions
of the paradox in total:
the original two (LFP and SFP), the attribute-based extensions (LAFP and
SAFP), the weighted extensions (LWFP and SWFP), and the weighted
attribute-based extensions (LWAFP and SWAFP). 
Given the lines of prior work that considered attributes and edge weights
separately, our framework aims
to bring
it all together both with common nomenclature for these variants,
and with the two general versions of the paradox---LEFP and
SEFP, equivalent to LWAFP and SWAFP---but we will find a midway stop
to talk about LWFP and SWFP useful. We aim for this paper to be a
one-stop venue for understanding the generalized phenomena from both a
theory and a data perspective. Note that despite \say{Friendship} in
the name, these algebraic concepts apply to networks across domains.
Inherent high clustering of social networks does have certain
implications, though, which we consider when looking at data.

When discussing the two original versions of the Friendship Paradox, we are looking at simple connected undirected finite graphs.
When working with the extensions that consider edge weights, we are looking at weighted undirected connected finite graphs, meaning that each edge weight is specified to be a positive integer (and all edge weights can still be $1$).
To make it explicit, all the graphs we consider still only have at most one edge between two nodes and do not have self-loops.

Many real-world networks are often not connected, but it is easy to see how the gaps can be calculated for the giant connected component (remove everything but the giant component) or for multiple components (iterate over nodes and their friends as usual). The only issue would arise for isolated nodes when looking at singular versions of the paradox: here, we couldn't find, say, the average degree of a node's friends because we'd have to be dividing by $0$. In practice, we usually disregard isolates when looking at both the singular and the list versions of the paradox.

To deal with real-world cases where the edge weights are positive but not necessarily integers, it is simplest to multiply each edge weight by a constant that would make each edge weight an integer (the least common multiple of all the edge weight denominators would work as such a constant). This would not affect the gap sign or whether the paradox holds. Here we do assume that each initial edge weight is a rational number or can be approximated by a rational number (any degree of precision would work). Our arguments for the extensions can be generalized to explicitly allow for any positive (including irrational) edge weights, but restricting edge weights to be positive integers allows for simpler definitions and proofs which we favor due to accessibility. That said, the numbered gap equations and the correlation rules work for positive real weights too.

Understanding the extensions and knowing how to use them for \say{simpler} cases (e.g. no arbitrary node attributes) and for more complicated cases (e.g. irrational edge weights, disconnected graphs) lets us cover many applications of the basic premise of \say{the friends' quantity is larger than the nodes' quantity} to undirected networks from any domain.

In Section~\ref{sec:fp}, we formalize the two original versions of the Friendship Paradox for undirected graphs and prove that they always hold. In Section~\ref{sec:wfp}, we allow for edge weights to be positive integers and prove that the two new versions also always hold. In Section~\ref{sec:efp}, we add arbitrary node attributes on top of that, formalize the two extensions and prove under what conditions they hold and fail. To illustrate how things work, we apply the two extensions to a 3-node network in Section~\ref{sec:example}; we both consider and disregard provided edge weights and arbitrary attributes and see how the extensions accommodate that, for a total of 8 gap values. (We cover the example very briefly in the main text, with much more detail in the Supplementary Information.) We look at simulation results in Section~\ref{sec:simulation} and at data results in Section~\ref{sec:data}. In the Supplementary Information, we consider FP for directed graphs and show that the 4 basic directed versions---list and singular, in-degree- and out-degree-based---do not hold in very simple cases.

In what follows, we define $n$ as the number of nodes in a graph; the nodes are indexed $1$ through $n$. The weight of the edge between nodes $i$ and $j$ is defined as $e_{ij}=e_{ji}$. (For completeness, $e_{ij}=e_{ji}=0$ if $i$ and $j$ aren't connected by an edge.) The number of neighbors $i$ has (the number of edges $i$ is part of) is denoted $d_i$ for both unweighted and weighted graphs and is known as the degree of node $i$, and $N(i)$ is the set of neighbors (or friends) of $i$. The sum of weights of edges that $i$ is part of, $\sum_{j \in N(i)} e_{ij}$, is denoted $w_i$ for both unweighted and weighted graphs and is known as the weighted degree of node $i$. Clearly, for unweighted graphs, $w_i = \sum_{j \in N(i)} e_{ij} = \sum_{j \in N(i)} 1 = d_i$. 

If all $d_i$ are equal in a graph, we call the graph regular. If all $w_i$ are equal, we call the graph \say{weighted-regular.} If the edge weights in a graph are not all $1$, the graph can be regular but not \say{weighted-regular} and vice versa, whereas if the edge weights are all $1$, regularity is of course equivalent to \say{weighted-regularity.} See the Supplementary Information for an example of a graph which is \say{weighted-regular} but not regular.

A numeric attribute of $i$ is denoted $a_i$. When we say that a graph \say{has no attributes} or that we are not considering attributes, we will take $a_i=w_i$ for each $i$. This will reduce to $a_i=d_i$ for unweighted graphs (the original Friendship Paradox), but will take non-trivial edge weights into account when they are present.

Finally, we will refer to the degree sequence, the weighted degree sequence, and the attribute sequence of a graph. These are lists of their associated values (e.g. the degree sequence is a list of degrees) indexed in the same way the nodes are.

\section{Friendship Paradox (FP)}
\label{sec:fp}
The Friendship Paradox is colloquially expressed as \say{your friends have more friends than you do} after the title of Scott Feld's paper \cite{feld1991your}. It's a statement about nodes' degrees and friends' degrees. There are two natural formalizations of the statement and we provide and prove both below.

\subsection{list version: LFP}
\label{sec:lfp}
\subsubsection{formal definition of the LFP gap}
\textit{List second-order degree} of node $i$ is the list of degrees of $i$'s friends; the length of such a list is $d_i$. To get the mean second-order degree (list version) of the graph as a whole, we concatenate (join) all such lists and take the mean of that; we will express that mean as a fraction. To get the numerator, we sum each list and then sum the sums, and the denominator is the sum of the lengths of all the lists, for a final value of $\frac{\sum_{i=1}^{n} \sum_{j \in N(i)} d_j}{\sum_{i=1}^{n} d_i}$ as the graph's mean second-order degree (list version). The LFP \textit{gap} is the difference between the mean second-order degree (list version) and the \say{mean first-order degree} (which is just mean degree) in a graph. That is expressed as:
\begin{equation}
\label{eqn:lfp}
g_{LFP} = \frac{\sum_{i=1}^{n} \sum_{j \in N(i)} d_j}{\sum_{i=1}^{n} d_i}-\frac{1}{n} \sum_{i=1}^{n} d_i
\end{equation}

In the expression above, nodes indexed $i$ are \say{seeds} taking stock of their friends' degrees. We can take a different approach and instead see how many times a friend's degree is featured in a seed's list second-order degree. A friend with degree $d_j$ is featured in $d_j$ seeds' calculations. The gap can thus be rewritten as:
$$g_{LFP} = \frac{\sum_{j=1}^{n} d^2_j}{\sum_{i=1}^{n} d_i}-\frac{1}{n} \sum_{i=1}^{n} d_i$$
Changing $j$ to $i$ and finding a common denominator, this is modified to:
$$g_{LFP} = \ddfrac{\sum_{i=1}^{n} d^2_i -\frac{1}{n} \left(\sum_{i=1}^{n} d_i \right) ^2}{\sum_{i=1}^{n} d_i}$$
We will say that LFP fails if $g_{LFP}$ is negative and holds otherwise.

\subsubsection{$g_{LFP}=0$ for regular graphs and positive for all non-regular graphs}
\label{sec:lfp_holds}
We claim that the gap is non-negative across our domain (with equality achieved for regular graphs): $g_{LFP} \geq 0$.

First note that the gap is indeed zero for regular graphs where every node has degree $d$, since the concatenated list of \say{list second-order degrees} is of the form $[d, d... d]$ (all nodes' friends have degree $d$) for a mean of $d$, and the list of nodes' degrees looks the same, with the only difference being that it may be shorter, for a mean of $d$. Now assume that the graph is non-regular. Since we are interested in the gap sign, let's look just at the numerator of $g_{LFP}$:
$$\text{numerator of } g_{LFP} = \sum_{i=1}^{n} d^2_i -\frac{1}{n} \left(\sum_{i=1}^{n} d_i \right) ^2$$
Reindex the degrees so that the following is true:
$$d_1 \geq d_2 \geq ... \geq d_n$$
Now we can see that the numerator of $g_{LFP}$ is positive by Chebyshev's sum inequality. The inequality is strict precisely because not all degrees are the same.

{\em Chebyshev's sum inequality. If $a_1 \geq a_2 \geq ... \geq a_n$ and $b_1 \geq b_2 \geq ... \geq b_n$, then $\frac{1}{n} \sum_{i=1}^{n} a_i b_i \geq \left( \frac{1}{n} \sum_{i=1}^{n} a_i \right) \left( \frac{1}{n} \sum_{i=1}^{n} b_i \right)$, with equality achieved if at least one of the sequences is constant} \cite{hardy1952inequalities}.

Recall that LFP fails if $g_{LFP}$ is negative and holds otherwise.
Thus LFP always holds.

Intuitively, the average friends' degree being higher than the average degree here is due to the fact that if node $i$'s degree is higher (say 10), $i$ has 10 friends, and those 10 friends each add $d_i=10$ to the second-order list, resulting in 10 being present 10 times, while a lower degree of 5 would only be present 5 times. So, nodes with higher-than-average degrees are over-represented in the friends' degree fraction (the mean second-order degree (list version)), and nodes with lower-than-average degrees are thus under-represented. Since each node contributes equally to the first-order calculation of the average degree in a graph, this makes the second-order mean higher. This intuitive argument, if helpful, is easiest to apply to this most basic version of the Friendship Paradox, LFP. In the other cases, we will just rely on algebra to see things.

\subsection{singular version: SFP}
\label{sec:sfp}
\subsubsection{formal definition of the SFP gap}
\textit{Singular second-order degree} of node $i$ is the mean of the degrees of $i$'s friends, expressed as a fraction $\frac{\sum_{j \in N(i)} d_j}{d_i}$ (or, the mean of $i$'s list second-order degree). The mean second-order degree (singular version) of the graph as a whole is the mean of such fractions for all nodes in a graph, expressed as $\frac{1}{n}\sum_{i=1}^{n} \left(\frac{1}{d_i} \sum_{j \in N(i)} d_j \right)$. The SFP \textit{gap} is the difference between the mean second-order degree (singular version) and the mean degree in a graph. That is expressed as:

\begin{equation}
\label{eqn:sfp}
g_{SFP} = \frac{1}{n}\sum_{i=1}^{n} \left(\frac{1}{d_i} \sum_{j \in N(i)} d_j \right) - \frac{1}{n}\sum_{i=1}^{n} d_i
\end{equation}
The term $\sum_{i=1}^{n} \left(\frac{1}{d_i} \sum_{j \in N(i)} d_j \right)$ deserves more attention. Note that if nodes $x$ and $y$ are connected by an edge, that edge contributes $\frac{d_x}{d_y}+\frac{d_y}{d_x}$ to this term. If $x$ and $y$ are not connected by an edge, they do not influence each other's second-order values. Let's define the set of edges of the graph as $E$. We can now rewrite the gap as:
$$g_{SFP} = \frac{1}{n} \sum_{(x,y)\in E} \left(\frac{d_x}{d_y}+\frac{d_y}{d_x} \right) - \frac{1}{n}\sum_{i=1}^{n} d_i $$
We will say that SFP fails if $g_{SFP}$ is negative and holds otherwise.

\subsubsection{$g_{SFP}=0$ for regular graphs and positive for all non-regular graphs}
\label{sec:sfp_holds}
We claim that the gap is non-negative across our domain (with equality achieved for regular graphs): $g_{SFP} \geq 0$.

First note that the gap is indeed zero for regular graphs where every node has degree $d$, since for each node, its singular second-order degree and its first-order degree are both equal to $d$, and so the second-order mean and the first-order mean are also both equal to $d$ for a difference of $0$. Now assume that the graph is non-regular.

Following Kramer et al. \cite{kramer2016multistep}, take the last definition of the gap:
$$g_{SFP} = \frac{1}{n} \sum_{(x,y)\in E} \left(\frac{d_x}{d_y}+\frac{d_y}{d_x} \right) - \frac{1}{n}\sum_{i=1}^{n} d_i $$
First note that for any $r \in \mathbb{R}$, $r + \frac{1}{r} \geq 2$ by the AM-GM inequality. The inequality states that $\frac{p+q}{2}\geq\sqrt{pq}$ for any real $p$ and $q$ with equality achieved if $p=q$, and we take $p=r$ and $q=\frac{1}{r}$.

Taking $r=\frac{d_x}{d_y}$ and $\frac{1}{r}=\frac{d_y}{d_x}$, we can say:
$$\frac{1}{n} \sum_{(x,y)\in E} \left(\frac{d_x}{d_y}+\frac{d_y}{d_x} \right) \geq \frac{1}{n} 2 \cdot |E|$$
Since the graph is non-regular and connected, for at least one edge $(x,y)$, $d_x\neq d_y$ and the AM-GM inequality is strict ($\left(\frac{d_x}{d_y}+\frac{d_y}{d_x} \right)>2$), so we can strengthen the overall inequality too:
$$\frac{1}{n} \sum_{(x,y)\in E} \left(\frac{d_x}{d_y}+\frac{d_y}{d_x} \right) > \frac{1}{n} 2 \cdot |E|$$
Note that the right-hand side is equal precisely to the mean degree. So,
$$\frac{1}{n} \sum_{(x,y)\in E} \left(\frac{d_x}{d_y}+\frac{d_y}{d_x} \right) > \frac{1}{n}\sum_{i=1}^{n} d_i$$
and
$$\frac{1}{n} \sum_{(x,y)\in E} \left(\frac{d_x}{d_y}+\frac{d_y}{d_x} \right) - \frac{1}{n}\sum_{i=1}^{n} d_i = g_{SFP} > 0$$

Recall that SFP fails if $g_{SFP}$ is negative and holds otherwise.
Thus SFP always holds.

\section{Weighted Friendship Paradox (WFP)}
\label{sec:wfp}
Here, we allow for edges to be weighted, each weight being a positive integer, and see how that affects first- and second-order values and our calculations. Conceptually, we will try to stay as close as possible to the original versions of the paradox.

In the original FP, we used degree as the node attribute. In the weighted version, we use \textit{weighted degree}, or the sum of the weights of the edges a node is part of. Also, if $e_{ij}>e_{ik}$, we want $j$ to matter more than $k$ in $i$'s second-order calculations. To make phrasing more straightforward and closer to that of the Extended FP (next section), we will refer to weighted degree as the attribute of $i$, but still denote it $w_i$.

The \textit{list second-order attribute} of node $i$ would still be the list of attributes of $i$'s friends, but now, if $j$ is a friend of $i$, $w_j$ would be present on the list $e_{ij}$ times and not necessarily once. The length of such a list for $i$ would be equal to the sum of the weights of the edges $i$ is part of, or $w_i$. The singular second-order attribute would still be the mean of the list second-order attribute.

So, we can define the LWFP gap, the difference between the mean second-order attribute (list version) and the mean attribute, as:
\begin{equation}
  g_{LWFP} = \frac{\sum_{i=1}^{n} \sum_{j \in N(i)} e_{ij} w_j}{\sum_{i=1}^{n} w_i}-\frac{1}{n} \sum_{i=1}^{n} w_i  
\end{equation}

And the SWFP gap, the difference between the mean second-order attribute (singular version) and the mean attribute, as:
\begin{equation}
g_{SWFP} = \frac{1}{n}\sum_{i=1}^{n} \left(\frac{1}{w_i} \sum_{j \in N(i)} e_{ij} w_j \right) - \frac{1}{n}\sum_{i=1}^{n} w_i
\end{equation}

\subsubsection{The WFP gaps are $0$ for ``weighted-regular'' graphs and positive otherwise}
\label{sec:wfp_holds}
We claim that both gaps are $0$ for \say{weighted-regular} graphs, i.e. graphs where all $w_i$ are equal, and positive otherwise, similar to the original FP.

To see why that is the case, it will actually be helpful to rewrite both gap formulas. Define $N_w(i)$ as the weighted neighborhood of $i$, which is the set of $i$'s friends except that each friend $j$ is represented $e_{ij}$ times. We can see that the gap expressions become:

$$g_{LWFP} = \frac{\sum_{i=1}^{n} \sum_{j \in N_w(i)} w_j}{\sum_{i=1}^{n} w_i}-\frac{1}{n} \sum_{i=1}^{n} w_i$$

$$g_{SWFP} = \frac{1}{n}\sum_{i=1}^{n} \left(\frac{1}{w_i} \sum_{j \in N_w(i)} w_j \right) - \frac{1}{n}\sum_{i=1}^{n} w_i $$

Now, compare these expressions to equations~\ref{eqn:lfp} and~\ref{eqn:sfp} for LFP and SFP gaps. The LWFP expression is almost identical to the LFP expression if we replace $w$ with $d$, and same for the singular version of WFP. But the weighted expressions deal with $N_w(i)$ instead of $N(i)$. To get around that, consider the following: we are only interested in seeds and friends, not the full structure of a network. For each seed $i$, for each of its friends $j$ such that $e_{ij}>1$, we can split edge $(i,j)$ into $e_{ij}$ new edges, and we can attach $j$ to the first edge, and copies of $j$ to the other edges. Now, the neighborhood we are looking at in WFP calculations is $N(i)$. We can still use $w_i$ (now equal to $d_i$ given that each edge weight is $1$) in the calculations, and we will still use each friend's attribute $w_j$. We have thus transformed our LWFP and SWFP equations into something that is algebraically identical to equations~\ref{eqn:lfp} and~\ref{eqn:sfp}. 
For FP, we showed that the gaps are $0$ if all $d_i$ are the same and positive otherwise. Since algebraically we get our LWFP and SWFP expressions by replacing $d$ with $w$ in the FP equations, we know that the LWFP and SWFP gaps are zero if all $w_i$ are the same (i.e. if the graph is \say{weighted-regular}), and positive otherwise.

\section{Extended Friendship Paradox (EFP)}
\label{sec:efp}
We were already talking about node attributes in Section~\ref{sec:wfp}. There, the attributes were equal to weighted degrees. Now, we can relax that and say that each attribute is arbitrary and denoted $a_i$. To get the gap formulations, we only need to replace the first-order quantities (that we are averaging) with $a_i$, but we will build the gap formulas from scratch to make things very clear. EFP is equivalent to the Weighted Attribute-based Friendship Paradox (WAFP). (To see how an attribute-based FP without edge weights is defined, see the Supplementary Information.) 

The attributes are indexed the same way as degrees and weighted degrees, and all these sequences have the same length, so we can precisely define a \textit{(weighted degree)-attribute correlation} which will be useful. Note though that if sequence $x$ consists of all the same values, the correlation between $x$ and $y$ (denoted $r_{x,y}$) is undefined, so we need to look at the case where all weighted degrees are equal and the case where all attributes are equal---meaning the two cases where the (weighted degree)-attribute correlation is undefined---separately. When the correlation between $x$ and $y$ is defined, it is equal to $r_{x,y} = \frac{\sum_{i=1}^n (x_i -\overline{x})(y_i - \overline{y})}{\sqrt{\sum_{i=1}^n (x_i - \overline{x})^2 \sum_{i=1}^n (y_i - \overline{y})^2 }}$.

\subsection{list version: LEFP}
\subsubsection{formal definition of the LEFP gap}
The \textit{list second-order attribute} of node $i$ is the list of attributes of its friends \textbf{weighted} by edge weight $e_{ij}$ for each friend $j$; in other words, for each friend $j$, $i$ takes $a_j$ and appends it to its own list $e_{ij}$ times instead of just once. The length of such a list is $\sum_{j \in N(i)} e_{ij}$, the weighted degree of $i$.

To get the mean second-order attribute (list version) of the graph as a whole, we take all such lists and take the mean of that; we will express that mean as a fraction. To get the numerator, we sum each list and then sum the sums, and the denominator is the sum of the lengths of all the lists. The LEFP \textit{gap} is the difference between the mean second-order attribute (list version) and the mean attribute in a graph. That is expressed as:
\begin{equation}
\label{eqn:lefp}
g_{LEFP} = \frac{\sum_{i=1}^{n} \sum_{j \in N(i)} e_{ij} a_j}{\sum_{i=1}^{n} \sum_{j \in N(i)} e_{ij}}-\frac{1}{n} \sum_{i=1}^{n} a_i
\end{equation}

Let's change $\sum_{j \in N(i)} e_{ij}$ in the denominator to $w_i$:
$$g_{LEFP} = \frac{\sum_{i=1}^{n} \sum_{j \in N(i)} e_{ij} a_j}{\sum_{i=1}^{n} w_i}-\frac{1}{n} \sum_{i=1}^{n} a_i$$

In the expression above, nodes indexed $i$ are \say{seeds} taking stock of their friends' degrees. We can take a different approach and instead see how many times a friend's degree is featured in a seed's list second-order attribute. A friend with weighted degree $w_j$ (and attribute $a_j$) is featured in seeds' calculations $w_j$ times (some of these times might be for the same seed). The gap can thus be rewritten as:

$$g_{LEFP} = \frac{\sum_{j=1}^{n} a_j w_j}{\sum_{i=1}^{n} w_i}-\frac{1}{n} \sum_{i=1}^{n} a_i$$

Changing $j$ to $i$ and finding a common denominator, this is modified to:
$$g_{LEFP} = \ddfrac{\sum_{i=1}^{n} a_i w_i -\frac{1}{n} \left(\sum_{i=1}^{n} a_i \right)\left(\sum_{i=1}^{n} w_i \right) }{\sum_{i=1}^{n} w_i}$$

We will say that LEFP fails if $g_{LEFP}$ is negative and holds otherwise.

\subsubsection{$g_{LEFP}=0$ for ``weighted-regular'' graphs and for graphs with no attribute variation}

First note that the gap is $0$ if the weighted degree sequence or the attribute sequence is constant. This is clear from a direct application of Chebyshev's sum inequality described above: if $a_1 = a_2 = ... = a_n$ or $w_1 = w_2 = ... = w_n$, then $\frac{1}{n} \sum_{i=1}^{n} a_i w_i = \left( \frac{1}{n} \sum_{i=1}^{n} a_i \right) \left( \frac{1}{n} \sum_{i=1}^{n} w_i \right)$ and the numerator of $g_{LEFP}$ is 0.

Now we can further assume that both weighted degrees and attributes have variation, meaning that the (weighted degree)-attribute correlation $r_{w,a}$ is defined.

\subsubsection{the sign of $r_{w,a}$ determines the sign of $g_{LEFP}$ for other graphs}
\label{sec:r_wa}
Note that if we subtract a constant $c$ from each node's attribute, the gap value doesn't change, since each element in the second-order list and the first-order list drops by $c$, and so the difference between the second-order mean and the first-order mean stays the same. Let's subtract the mean of the attribute sequence from each attribute. Now the attribute sequence mean, which we will denote $\overline{a}$, is $0$, and the gap didn't change.

Taking $\overline{a}=0$, our gap formula becomes:
$$g_{LEFP} = \frac{1}{\sum_{i=1}^{n} w_i} \sum_{i=1}^{n} w_i a_i $$
with a positive coefficient $\frac{1}{\sum_{i=1}^{n} w_i}$.

The (weighted degree)-attribute correlation $r_{w,a}$, given $\overline{a}=0$, is:

$$r_{w,a} (\overline{a}=0) = \frac{\sum_{i=1}^n a_i (w_i - \overline{w})}{\sqrt{\sum_{i=1}^n a_i^2 \sum_{i=1}^n (w_i - \overline{w})^2 }} $$
with a positive denominator.

The sign of the correlation is thus determined by the sign of the numerator:
$$ \sum_{i=1}^n a_i (w_i - \overline{w})$$
From here, we get
$$ \sum_{i=1}^n a_i w_i - \sum_{i=1}^n a_i \overline{w} = 
\sum_{i=1}^n a_i w_i - \overline{w} \sum_{i=1}^n a_i =
\sum_{i=1}^n a_i w_i - \overline{w} 0 =
\sum_{i=1}^n w_i a_i
$$
So, the sign of $r_{w,a}$ is equal to the sign of $\sum_{i=1}^n w_i a_i$. Since $g_{LEFP}$ is $\sum_{i=1}^n w_i a_i$ multiplied by a positive number, the sign of $r_{w,a}$ is equal to the sign of $g_{LEFP}$.

Recall that LEFP fails if $g_{LEFP}$ is negative and holds otherwise.
LEFP holds in cases when $r_{w,a}$ is undefined (when the weighted degrees are all the same and/or the attributes are all the same). If $r_{w,a}$ is defined, LEFP fails if $r_{w,a}<0$ and holds otherwise. Note that the gap being $0$ does not make a statement about a graph's \say{weighted-regularity} (or regularity like it does in the case of the original FP versions), since the LEFP gap is also $0$ when $r_{w,a}=0$.

\subsubsection{LEFP when not considering attributes}
\label{sec:lefp_lwfp}
If we don't have arbitrary attributes and want to implement LWFP via LEFP, we may want to replace attributes with degrees in equation~\ref{eqn:lefp}, since degrees are used in FP. That would be incorrect. Instead, we need to replace each attribute $a_i$ with the sum of weights of edges that $i$ is part of, or $i$'s weighted degree $w_i$. See Section~\ref{sec:wfp} for details on how WFP is set up. Using regular degrees as first-order values but using edge weights in second-order-value calculations is not LWFP or LFP which do not fail---it is a case of LWAFP with degrees as attributes, and the gap sign is equal to the sign of $r_{w,d}$. See the Supplementary Information for an example of a graph where this correlation is negative and LWAFP fails.

\subsubsection{LEFP failure when considering attributes}
\label{sec:lefp_fails}
When we don't consider attributes and instead use weighted degrees as first-order values, the correlation $r_{w,a}$ reduces to $r_{w,w}$ and the paradox doesn't fail. If we do have arbitrary attributes, we can estimate how often the list version of the extended paradox may fail. Imagine you have a graph with a fixed weighted degree sequence $w$. Take and an arbitrary attribute assignment $a$. Suppose $r_{w,a}$ is positive. Here, LEFP holds. Now flip the sign of all the attributes. This flips the sign of the correlation and the gap---LEFP fails. If the initial $r_{w,a}$ is negative, the flipped version leads to $r_{w,a}$ being positive. While zero $r_{w,a}$ correlation is possible too, and the opposite of that attribute sequence would also lead to a correlation of zero (and to both gaps being zero meaning the paradox holds), and it's possible to have constant attribute sequences or weighted degree sequences for which the gap is zero (and so the flipped-attributes gap is zero too), these cases are rare. Thus this thought experiment shows that LEFP, when considering arbitrary attributes, may fail in around 50\% cases in theory. When we deal with real-world data, this number may be very different since the attributes may be positively or negatively correlated with weighted degrees.

\subsection{singular version: SEFP}
\subsubsection{formal definition of the SEFP gap}
The easiest way to get the \textit{singular second-order attribute} of node $i$ is to take the mean of the \textit{list second-order attribute} of node $i$, which is a list of node $i$'s friends attributes where each $a_j$ is represented $e_{ij}$ times. The length of this list is equal to $\sum_{j \in N(i)} e_{ij}$ and its mean is expressed as $\frac{\sum_{j \in N(i)} e_{ij} a_j}{\sum_{j \in N(i)} e_{ij}}$. The mean second-order attribute (singular version) of the graph as a whole is the mean of such fractions for all nodes in a graph, expressed as $\frac{1}{n}\sum_{i=1}^{n} \left(\frac{1}{\sum_{j \in N(i)} e_{ij}} \sum_{j \in N(i)} e_{ij} a_j \right)$. The SEFP \textit{gap} is the difference between the mean second-order attribute (singular version) and the mean attribute in a graph. That is expressed as:
\begin{equation}
\label{eqn:sefp}
g_{SEFP} = \frac{1}{n}\sum_{i=1}^{n} \left(\frac{1}{\sum_{j \in N(i)} e_{ij}} \sum_{j \in N(i)} e_{ij} a_j \right) - \frac{1}{n}\sum_{i=1}^{n} a_i
\end{equation}
Let's change $\sum_{j \in N(i)} e_{ij}$ to $w_i$:
$$g_{SEFP} = \frac{1}{n}\sum_{i=1}^{n} \left(\frac{1}{w_i} \sum_{j \in N(i)} e_{ij} a_j \right) - \frac{1}{n}\sum_{i=1}^{n} a_i$$

In the expression above, index $i$ refers to nodes (\textit{seeds}) and index $j$ to $i$'s friends, and we look at the seeds' calculations of their second-order attributes. But we can also do the opposite and see how each node $j$'s attribute features in its friends' second-order values. $a_j$'s coefficient in the second-order mean is equal to $\frac{1}{n}$ times a cumbersome quantity, the weighted sum of $j$'s friends' reciprocal weighted degrees: 
$$g_{SEFP} = \frac{1}{n}\sum_{j=1}^{n} a_j \left( \sum_{k \in N(j)} \frac{e_{jk}}{w_k} \right) - \frac{1}{n}\sum_{j=1}^{n} a_j$$

The equation above may be the one that works best when we actually want to compute the gap size.

We call $\sum_{j \in N(i)} \frac{e_{ij}}{w_j}$, the \textbf{weighted} sum of $i$'s friends' reciprocal weighted degrees, $\gamma_i$ (gamma). (For an unweighted graph, a corresponding quantity encountered in literature is delta, the sum of $i$'s friends' reciprocal degrees, $\sum_{j \in N(i)} \frac{1}{d_j}=\delta_i$~\cite{evtushenko2023node}. Gamma reduces to delta if all edge weights are $1$.)

We can now rewrite the previous expression as:
$$g_{SEFP} = \frac{1}{n}\sum_{j=1}^{n} \gamma_j a_j - \frac{1}{n}\sum_{j=1}^{n} a_j$$

Let's reindex $j$ to $i$ for a final formulation:
\begin{equation}
\label{eqn:sefp2}
g_{SEFP} = \frac{1}{n}\sum_{i=1}^{n} \gamma_i a_i - \frac{1}{n}\sum_{i=1}^{n} a_i
\end{equation}

We will say that SEFP fails if $g_{SEFP}$ is negative and holds otherwise.

\subsubsection{$g_{SEFP}=0$ for ``weighted-regular'' graphs and for graphs with no attribute variation}

If all weighted degrees are the same and equal to $w$, each $\gamma_i$ is equal to $\sum_{j \in N(i)} \frac{e_{ij}}{w_j} = \sum_{j \in N(i)} \frac{e_{ij}}{w} = \frac{1}{w}\sum_{j \in N(i)} e_{ij} =  \frac{1}{w} w_i  = \frac{w}{w} = 1$. Looking at Equation~\ref{eqn:sefp2}, this makes the gap $0$.

If all attributes the same and equal to $a$, using Equation~\ref{eqn:sefp2}, we have 
$$g_{SEFP} = \frac{1}{n}\sum_{i=1}^{n} \gamma_i a - \frac{1}{n}\sum_{i=1}^{n} a =\frac{1}{n} a (\sum_{i=1}^{n} \gamma_i - n)$$
The sign of the gap is equal to the sign of $\sum_{i=1}^{n} \gamma_i - n$. The sum of gammas $\sum_{i=1}^{n} \gamma_i$ is equal to $\sum_{i=1}^{n} \sum_{j \in N(i)} \frac{e_{ij}}{w_j}$. In this sum of sums, the final coefficient of term $\frac{1}{w_j}$ is $w_j$, since each of $j$'s friends $k$ uses $\frac{1}{w_j}$ in their calculation of $\gamma_k$ (their gamma) $e_{jk}$ times, and the sum of $e_{jk}$ over all such nodes $k$ is $w_j$. Since this argument applies to every $j$'s reciprocal weighted degree $\frac{1}{w_j}$, the sum of all gammas is equal to $\sum_{j=1}^{n} w_j \frac{1}{w_j} = \sum_{j=1}^{n} 1 = n$. Thus $\sum_{j=1}^{n} \gamma_j - n$ is equal to $0$, the sign of the SEFP gap is \say{none} and the gap itself is $0$.

Note also that if a graph is not \say{weighted-regular} (a case we just looked at), we have variation in gammas. We provide the proof in the Supplementary Information. Once we have established that there is variation in gammas, we can further assume that $r_{\gamma,a}$ is defined.

\subsubsection{the sign of $r_{\gamma,a}$ determines the sign of $g_{SEFP}$ for other graphs}
\label{sec:r_gammaa}
We saw before that subtracting the mean of the attribute sequence from each attribute makes the new attribute sequence mean $0$ and doesn't change the gap.
Taking $\overline{a}=0$, our gap formula becomes:
$$ g_{SEFP} = \frac{1}{n}\sum_{i=1}^{n} \gamma_i a_i $$
with a positive coefficient $\frac{1}{n}$.

The correlation $r_{\gamma,a}$, given $\overline{a}=0$, is:
$$r_{\gamma,a} (\overline{a}=0) = \frac{\sum_{i=1}^n a_i (\gamma_i - \overline{\gamma})}{\sqrt{\sum_{i=1}^n a_i^2 \sum_{i=1}^n (\gamma_i - \overline{\gamma})^2 }} $$
with a positive denominator.

The sign of the correlation is thus determined by the sign of the numerator:
$$ \sum_{i=1}^n a_i (\gamma_i - \overline{\gamma})$$
From here, we get
$$ \sum_{i=1}^n a_i \gamma_i - \sum_{i=1}^n a_i \overline{\gamma} = 
\sum_{i=1}^n a_i \gamma_i - \overline{\gamma} \sum_{i=1}^n a_i =
\sum_{i=1}^n a_i \gamma_i - \overline{\gamma} 0 =
\sum_{i=1}^n \gamma_i a_i
$$
So, the sign of $r_{\gamma,a}$ is equal to the sign of $\sum_{i=1}^n \gamma_i a_i$. Since $g_{SEFP}$ is $\sum_{i=1}^n \gamma_i a_i$ multiplied by a positive number, the sign of $r_{\gamma,a}$ is equal to the sign of $g_{SEFP}$.

Recall that SEFP fails if $g_{SEFP}$ is negative and holds otherwise.
SEFP holds in cases when $r_{\gamma,a}$ is undefined (when the weighted degrees are all the same and/or the attributes are all the same). If $r_{\gamma,a}$ is defined, SEFP fails if $r_{\gamma,a}<0$ and holds otherwise. Note that the gap being $0$ does not make a statement about a graph's \say{weighted-regularity} (or regularity like it does in the case of the original FP versions), since the SEFP gap is also $0$ when when $r_{\gamma,a}=0$.

\subsubsection{SEFP when not considering attributes}
If we don't have arbitrary attributes and want to implement SWFP via SEFP, we, similarly to the logic in Section~\ref{sec:lefp_lwfp}, need to use weighted degrees, and not simply degrees, in place of attributes. See the Supplementary Information for an example where SWAFP with degrees as attributes fails, which is equivalent to $r_{\gamma,d}$ being negative.

\subsubsection{SEFP failure when considering attributes}
\label{sec:sefp_fails}
When when we don't consider attributes, SEFP reduces to SWFP which doesn't fail according to the logic from Section~\ref{sec:wfp_holds}. If we do have arbitrary attributes, we can estimate how often the singular version of the extended paradox may fail similar to how we did it for LEFP in Section~\ref{sec:lefp_fails}. Imagine you have a graph with a fixed weighted degree sequence $w$ and thus a fixed gamma sequence $\gamma$. Take an arbitrary attribute assignment $a$. Suppose $r_{\gamma,a}$ is positive. Here, SEFP holds. Now flip the sign of all the attributes. This flips the sign of the correlation and the gap---SEFP fails. If the initial $r_{\gamma,a}$ is negative, the flipped version leads to $r_{\gamma,a}$ being positive. While zero $r_{\gamma,a}$ correlation is possible too, and the opposite of that attribute sequence would also lead to a correlation of zero (and to both gaps being zero meaning the paradox holds), and it's possible to have constant attribute sequences or weighted degree sequences for which the gap is zero (and so the flipped-attributes gap is zero too), these cases are rare. Thus this thought experiment shows that SEFP, when considering arbitrary attributes, may fail in around 50\% cases in theory. When we deal with real-world data, this number may be very different since the attributes may be positively or negatively correlated with gammas.

\subsection{How reductions work}
Now that we have defined LEFP and SEFP, we want to be able to apply it to all kinds of graphs, whether or not they have edge weights or arbitrary attributes defined.

It is easiest to use Equations~\ref{eqn:lefp} and~\ref{eqn:sefp} for gap calculations.

Most importantly, if no arbitrary attributes are present, each attribute $a_i$ in the equations should be equal to the sum of weights of edges that $i$ is part of, which is equal to $w_i$ and not $d_i$. ($w_i$ will reduce to $d_i$ for the case of no edge weights.)

When no edge weights are defined, simply define them all to be $1$.

It's also interesting to see what happens to the two correlation-based rules when we don't consider arbitrary attributes or edge weights. 

We know that
$r_{w,a}$ determines the sign of $g_{LWAFP}$ and 
$r_{\gamma,a}$ determines the sign of $g_{SWAFP}$. 

If we don't consider edge weights, we see that 
$r_{d,a}$ determines the sign of $g_{LAFP}$ and
$r_{\delta,a}$ determines the sign of $g_{SAFP}$. (Delta of $i$ is the sum of the reciprocal degrees of $i$'s friends, $\sum_{j \in N(i)} \frac{1}{d_j}$.)

If we don't consider arbitrary attributes, we see that 
$r_{w,w}$ determines the sign of $g_{LWFP}$ and
$r_{\gamma,w}$ determines the sign of $g_{SWFP}$. 
Since $r_{w,w}=1$ every time it's defined, and since LWFP holds whenever $r_{w,w}$ is not defined (when the graph is \say{weighted-regular}), we see that LWFP always holds (which we also know from algebra in Section~\ref{sec:wfp_holds}). Conversely, since we know from that same section that SWFP always holds, we see that $r_{\gamma,w}$, when defined, is always positive, which is a non-trivial result.

If we don't consider edge weights or arbitrary attributes, we see that
$r_{d,d}$ determines the sign of $g_{LFP}$ and
$r_{\delta,d}$ determines the sign of $g_{SFP}$.
Like above, this tells us that LFP always holds and that $r_{\delta,d}$, when defined, is positive.

Because weighted degrees can be defined virtually independently from degrees (via assigning arbitrary weights to edges), there is no direct relationship between $r_{d,a}$ and $r_{w,a}$, or $r_{\delta,a}$ and $r_{\gamma,a}$. Similarly, $r_{d,a}$ being positive ($g_{LAFP}$ being positive) doesn't necessarily imply that $r_{\delta,a}$ is positive ($g_{SAFP}$ being positive). The same can be said for the weighted FP counterparts of these phenomena. We did show in a previous paper that $r_{d,\delta}$ being $1$ would naturally create a sign dependency between $r_{d,a}$ and $r_{\delta,a}$, and in fact they would be equal \cite{evtushenko2023node}. We showed that the graphs for which $r_{d,\delta}$ was $1$ were difficult to describe in any other way, for example in terms of \say{looks.} A similar statement can be made for the WAFP versions, with $r_{w,\gamma}=1$ implying that $r_{w,a}$ and $r_{\gamma,a}$ have the same sign and are in fact equal.

\section{Applying the extensions to one small network}
\label{sec:example}
It may be helpful to see how things work on a very small example in Figure~\ref{fig:example}:
\begin{figure}[H]
\centering
\includegraphics[width=0.4\textwidth]{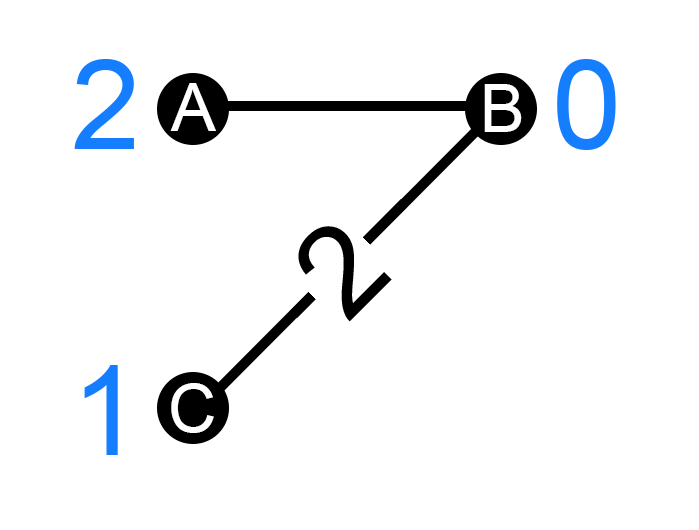}
\caption{This is a weighted undirected graph $G$ on 3 nodes. The weight of each edge, when not 1, is listed on the edge. The nodes are labeled $A$ through $C$. Each node has a numeric attribute $a_i$ associated with it, and it's listed in blue next to each node.
\label{fig:example}}
\end{figure}
We can look at all the data presented, or we can ignore edge weights and/or node attributes. We want to apply an extension (LEFP or SEFP) in every case. To keep track of what we are disregarding, we will use the names Friendship Paradox (an extension applied to the graph without weights or attributes), Attribute-based Friendship Paradox (an extension applied to the graph without weights), Weighted Friendship Paradox (an extension applied to the graph without attributes) and Weighted Attribute-based Friendship Paradox (EFP) (an extension applied to the graph as is).

We saw in Sections~\ref{sec:fp} and \ref{sec:wfp} that FP and WFP always hold. When considering attributes, though (whether or not we also consider edge weights), the paradox may fail depending on the signs of the correlations specified in Section~\ref{sec:efp}. To illustrate this, in our example, the four versions that can fail do fail, but this is clearly not always the case. 

We only provide the gap sizes (and signs) in this section---for an extended argument, see the Supplementary Information.

$g_{LFP}=\frac{1}{6}$, $g_{SFP}=\frac{1}{3}$; $g_{LAFP}=-\frac{1}{4}$ which follows $r_{d,a}$ being negative at $-0.866$; $g_{SAFP}=-\frac{1}{2}$ which follows $r_{\delta,a}$ being negative at $-0.866$; $g_{LWFP}=\frac{1}{3}$, $g_{SWFP}=\frac{5}{9}$; $g_{LWAFP}=-\frac{1}{3}$ which follows $r_{w,a}$ being negative at $-1$; $g_{SWAFP}=-\frac{5}{9}$ which follows $r_{\gamma,a}$ being negative at $-0.944$.

\section{Simulation}
\label{sec:simulation}
Since we are not interested in gap sizes and since the non-attribute-based versions of the paradox do not fail, we are mainly considering attribute-based versions here: WAFP (EFP) where we also look at edge weights, and AFP where we don't look at edge weights. We are interested in how increasing the expected value of an attribute-based correlation (say, $r_{w,a}$) affects how often the associated version of the paradox (in this case, LWAFP) fails. While degrees, weighted degrees, deltas and gammas aren't necessarily highly correlated, raising/lowering $r_{d,a}$ \textbf{in our construction} affects $r_{w,a}$, $r_{\delta,a}$ and $r_{\gamma,a}$ similarly, so we will focus on changing $r_{d,a}$ since degree is the most intuitive concept out of the four and changing attributes in one way is simpler than in four ways. 

This simulation is not meant to be exhaustive, but merely illustrative. For each calculation, we will also confirm that the sign of the correlation we are looking at matches the sign of its associated gap.

We use a $G_{n,p}$ random graph generator with graph size $n=1000$ and edge probability $p=\frac{1}{50}$ ($p=\frac{1}{50}$ was initially selected to get sparser---more computationally efficient---graphs that are still often connected for $n=1000$, but we later discovered another good reason to settle on this, which we mention below). Because our domain is connected graphs, we discard disconnected graphs produced by $G_{n,p}$. (We would also discard regular graphs and graphs that end up being \say{weighted-regular} once we assign edge weights, but in practice, we found neither disconnected, nor connected regular, nor connected \say{weighted-regular} graphs and so did not need to discard them.)

We think of an edge weight as the strength of a connection between two nodes. While it's plausible that if one person has strong ties (edge weight $>>1$) with two people, the two people also have a tie stronger than $1$ (an extension of the strong triadic closure property~\cite{granovetter1973strength}), to keep things simple, we assign weights to existing network edges independently and distribute them uniformly between $1$ and 10 (restricting to integers).

We do $1000$ runs of the simulation. A run includes creating $n=1000$ nodes, labelling them $1$ through $1000$, determining which edges exist, confirming the graph is connected and not regular, determining the weight of each edge, confirming the graph is not \say{weighted-regular,} and drawing an attribute sequence $a$ of size $1000$ from a standard normal distribution with mean $0$ and s.d. $1$ ($\mathcal{N}(0,\,1)$). For each condition listed below, we alter that sequence to make $r_{d,a}$ smaller or larger, and calculate the four gaps ($g_{LAFP}$, $g_{SAFP}$, $g_{LWAFP}$, and $g_{SWAFP}$) and the four associated correlations ($r_{d,a}$, $r_{\delta,a}$, $r_{w,a}$, and $r_{\gamma,a}$). To summarize things across 1000 cases, for each condition we keep track of the $4$ mean correlations and the proportion of cases where each of the paradoxes fails.

\subsection{Simulation conditions}

In condition $0$, each of our attributes remains as is (drawn from $\mathcal{N}(0,\,1)$). The expected degree-attribute correlation is $0$ and so we expect LAFP to fail in around 50\% of the $1000$ cases. (Recall that it's the LAFP gap sign that is equal to the $r_{d,a}$ sign.)

In positive-numbered conditions $j$ ($j \in \{1,2,...100\}$), attribute $a_i$ is the sum of the original number drawn from $\mathcal{N}(0,\,1)$ and $\frac{j}{100} d_i$. As we move towards higher-indexed conditions, the degree-attribute correlation $r_{d,a}$ increases in expectation. We expect LAFP to fail less often.

In negative-numbered conditions $j$ ($j \in \{-1,-2,...-100\}$), attribute $a_i$ is similarly the sum of the original number drawn from $\mathcal{N}(0,\,1)$ and $\frac{j}{100} d_i$. Here, as we move towards lower-indexed conditions, $r_{d,a}$ decreases in expectation. We expect LAFP to fail more often.

Making $r_{d,a}$ higher or lower will also affect $r_{\delta,a}$, $r_{w,a}$, and $r_{\gamma,a}$. All four parameters are highly correlated in our construction. $r_{d,\delta}$ has mean $0.973$ and s.d. $0.0013$ across the $1000$ networks, $r_{w,\gamma}$ has mean $0.966$ and s.d. $0.0017$, and $r_{d,w}$ has mean $0.884$ and s.d. $0.006$.

(Notably, when using $G_{n,p}$, the degree-delta correlation $r_{d,\delta}$ depends on the choice of $n$ and $p$. All things being equal, higher $n$ (bigger graph) raises $r_{d,\delta}$, and all things being equal, higher $p$ (denser graph) raises $r_{d,\delta}$. But raising either makes correlation and gap computations take longer. We have found $n=1000$ and $p=\frac{1}{50}$ to have a good balance of high $r_{d,\delta}$ (and high $r_{w,\gamma}$) and reasonable computation times).

\subsection{Simulation results}
The first thing to note is that all $1000$ networks we generated were connected and not regular or \say{weighted-regular,} so we didn't have to discard anything. Secondly, across the $201$ conditions, we never had a $0$ gap (as a reminder, the gap would be $0$ if its associated correlation was precisely $0$, if the graph was regular (AFP) or \say{weighted-regular} (WAFP), or if all the attributes were the same). We also never encountered a scenario where the gap sign was different from the sign of its associated correlation which confirms our theoretical findings.

\begin{figure}[H]
\centering
\includegraphics[width=0.7\textwidth]{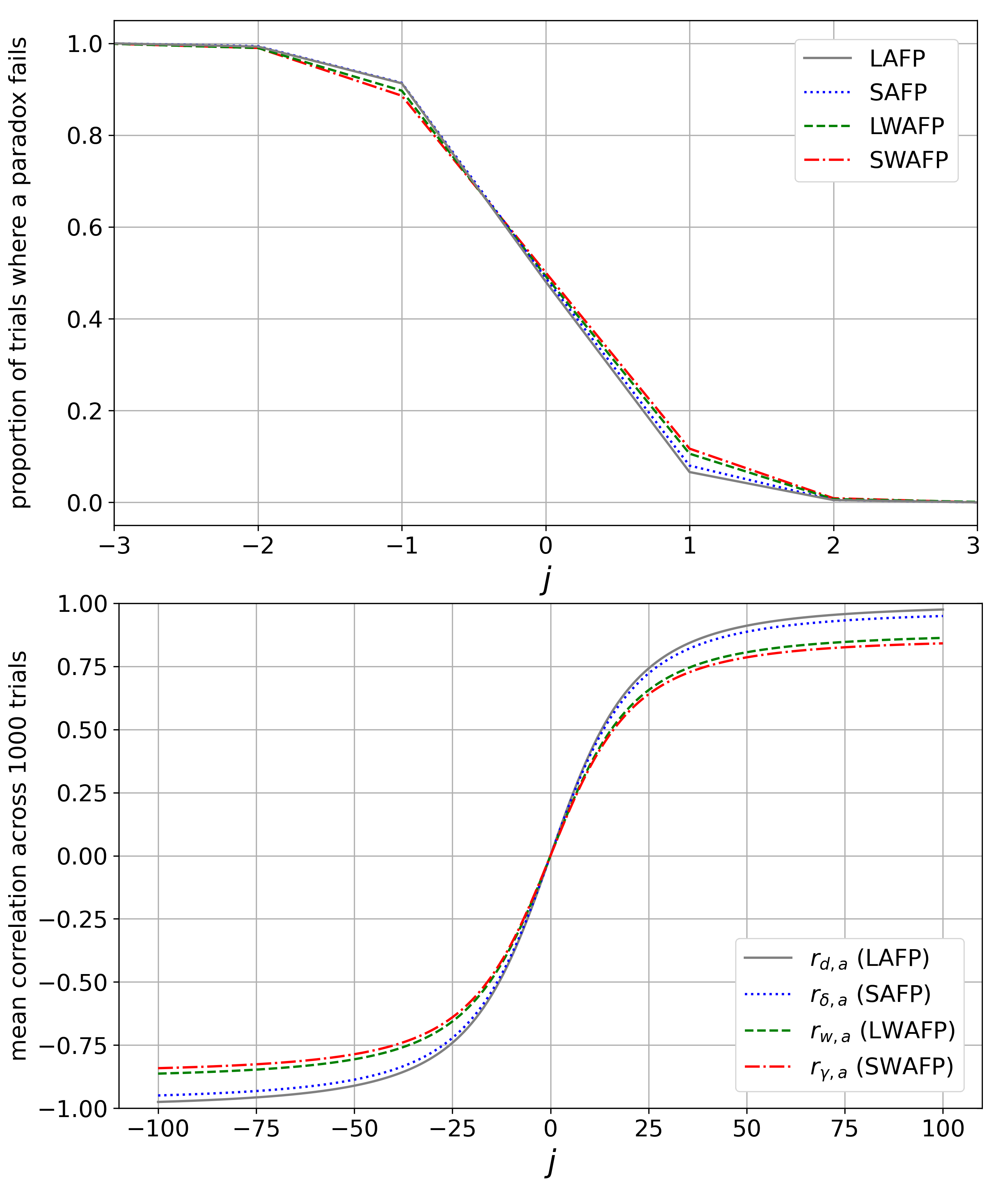}
\caption{The proportion of cases in which each paradox fails resembles a step function with $j$ as the argument---that is why in the top panel we restrict $j$ to $[-3,3]$ so the shift is more apparent. Since positive $j$ implies a positive degree-attribute correlation which, in our construction, is tightly linked to other correlations being positive, and correlation needs to be just slightly greater than or equal to $0$ for an attribute-based version of the paradox to hold, it makes sense that the gaps become positive and the proportion of failure drops to 0 in a step-like fashion. A contributing factor is that given our construction, the standard deviation of the $1000$ correlations for each condition is low ($<0.035$ and highest at $j=0$). See the Supplementary Information for a plot of the standard deviation. (Note also that in the top panel, the minute details of the lines' behavior around 0 are due to randomness which is highest at $j=0$. Here, the lines seemingly all cross around $-0.5$, but that isn't the case in each run of the simulation. We expect the proportion of failure to be $0.5$ at $j=0$ for each line, but it's not guaranteed to be precisely that in a simulation.)
\label{fig:simulation_results}}
\end{figure}

Overall, our predictions that 
\begin{itemize}
    \item as we increase the expected degree-attribute correlation, LAFP, SAFP, LWAFP and SWAFP fail less often
    \item as we decrease the expected degree-attribute correlation, the opposite happens
\end{itemize}
were correct.
See the results in Figure~\ref{fig:simulation_results}.

\begin{figure}[H]
\centering
\includegraphics[width=\textwidth]{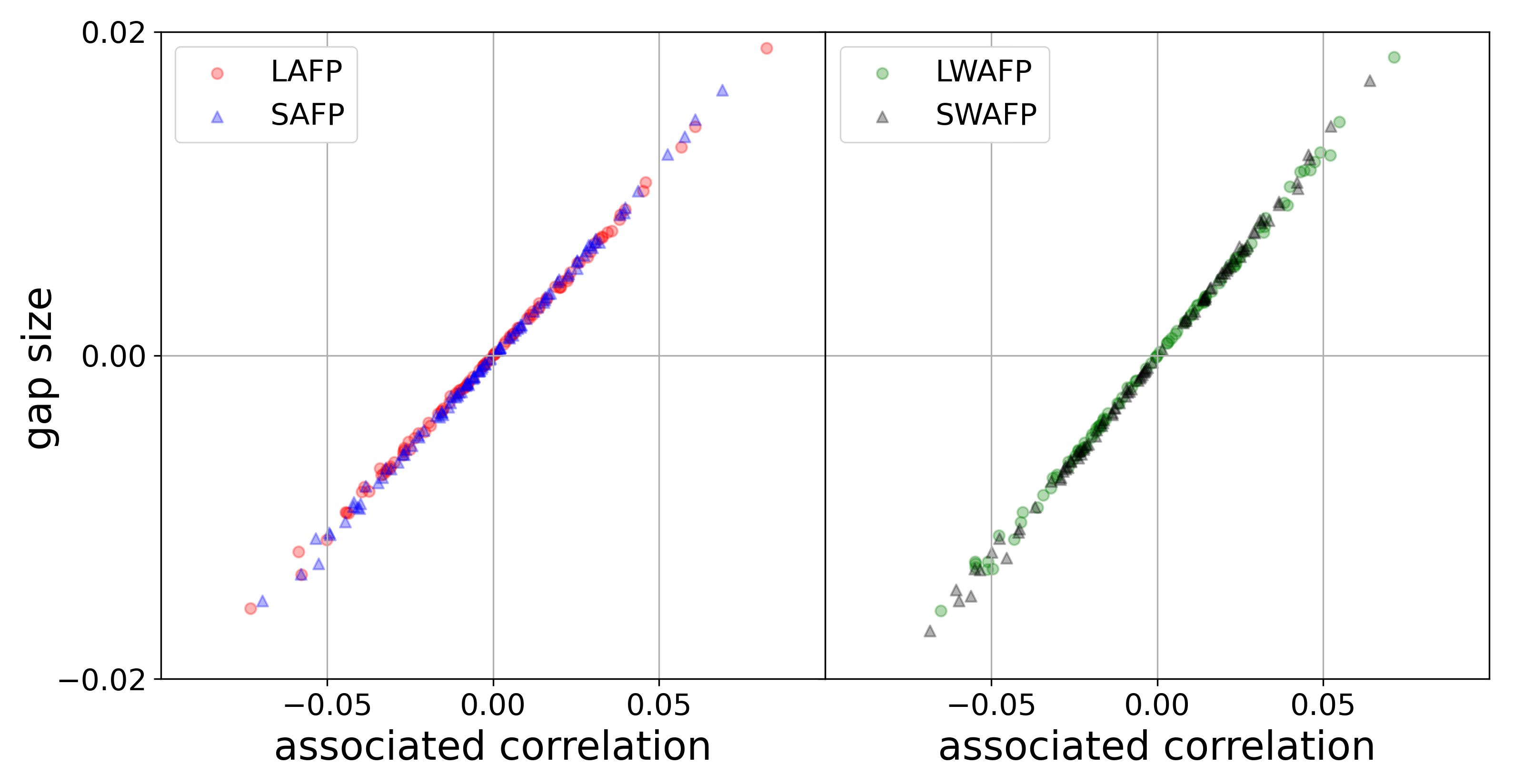}
\caption{We created a standard normal attribute sequence for each of the $1000$ $G_{1000,\frac{1}{50}}$ networks and found $r_{d,a}$, $r_{\delta,a}$, $r_{w,a}$ and $r_{\gamma,a}$, and the LAFP, SAFP, LWAFP and SWAFP gap sizes for each. The LAFP gap signs follow the $r_{d,a}$ signs, the SAFP gap signs follow the $r_{\delta,a}$ signs, the LWAFP gap signs follow the $r_{w,a}$ signs, and SWAFP gap signs follow the $r_{\gamma,a}$ signs. Furthermore, the correlation between the $x$-axis quantity and the $y$-axis quantity is $0.9994$ for all four pairs (the exact values are slightly different). But the correlation may not be as high for less symmetric cases such as real-world data. While we aren't interested in gap sizes on their own, it is interesting to see a strong linear relationship between them and their associated correlations \textbf{in this specific case}. Note: for illustration purposes we only plot the results for $100$ networks out of $1000$. 
\label{fig:j0}}
\end{figure}

The correlation spread is highest at $j=0$ when the attributes are independent of the degrees: the standard deviation is around $0.032$ for $r_{d,a}$, $r_{\delta,a}$, $r_{w,a}$ and $r_{\gamma,a}$. It's interesting to see the relationship between correlation values and gap values for that specific example (where attributes are distributed normally), since the correlations (plotted on the $x$-axis) would both vary and have different signs. See the results of that in Figure~\ref{fig:j0} (we plot $100$ points of each kind instead of $1000$). All $4000$ LFP, SFP, LWFP and SWFP gaps are positive. LAFP fails in $503$ out of $1000$ cases (the gap is negative), SAFP fails in $511$, LWAFP fails in $484$ and SWAFP fails in $490$. None of the gaps are zero. This confirms our logic in~Sections~\ref{sec:lefp_fails} and~\ref{sec:sefp_fails} where we concluded that attribute-based versions fail around 50\% of the time given random attribute assignment.

\section{Data}
\label{sec:data}

Our simulation results are very straightforward because our construction is very simple. Real-world data is messier and it's interesting to see how gap sizes and their associated correlations behave in those cases. Here again, we are only interested in the gap sizes insofar as they relate to correlations and so focus on LAFP, SAFP, LWAFP, and SWAFP, but we'd also like to report that all the LFP, SFP, LWFP, and SWFP gaps we find in the data discussed below are positive.

\subsection{Facebook100}
We have found the Facebook100 dataset useful in the past (see \cite{traud2012social,altenburger2018monophily} for details on this data). Each of the $100$ networks is a 2005 snapshot of Facebook friendships within a US university. Each network is connected; the average network size is $12083.16$; each network has an average degree and the average average degree across the networks is $76.78$. Nodes are people. They have attributes such as gender (\say{female,} \say{male,} or \say{not reported}), graduation year (including \say{not reported}) and other (anynonymized) attributes such as high school and major. The average proportion of nodes with no reported gender is $0.0767$ across the $100$ networks. The average proportion of nodes with no reported year of graduation is $0.137$ across the $100$ networks. Facebook100 is useful because it gives us a chance to create something like Figure~\ref{fig:j0}, but for real-world data where the graph size and other parameters change, but the context stays the same.

The only thing Facebook100 does not have is edge weights since Facebook friendships are not weighted. Using the principle of homophily, we can suppose that people with the same graduation year may have stronger ties and assign to edges between two such people a weight of $2$. Any other two nodes connected by an edge (including pairs where both nodes have unreported year of graduation) would have an edge of weight $1$. Across the $100$ networks, the average proportion of edges that get a weight of $2$ is $0.457$.

We'd also like to create a numeric attribute for each node, since the existing attributes are not reported for all the nodes and/or are categorical in nature. We will use \say{prop\_own,} the proportion of a node's friends that share its gender value (including \say{not reported}). Each network would have an average attribute and the average average attribute across the $100$ networks is $0.496$.

\begin{figure}[H]
\centering
\includegraphics[width=\textwidth]{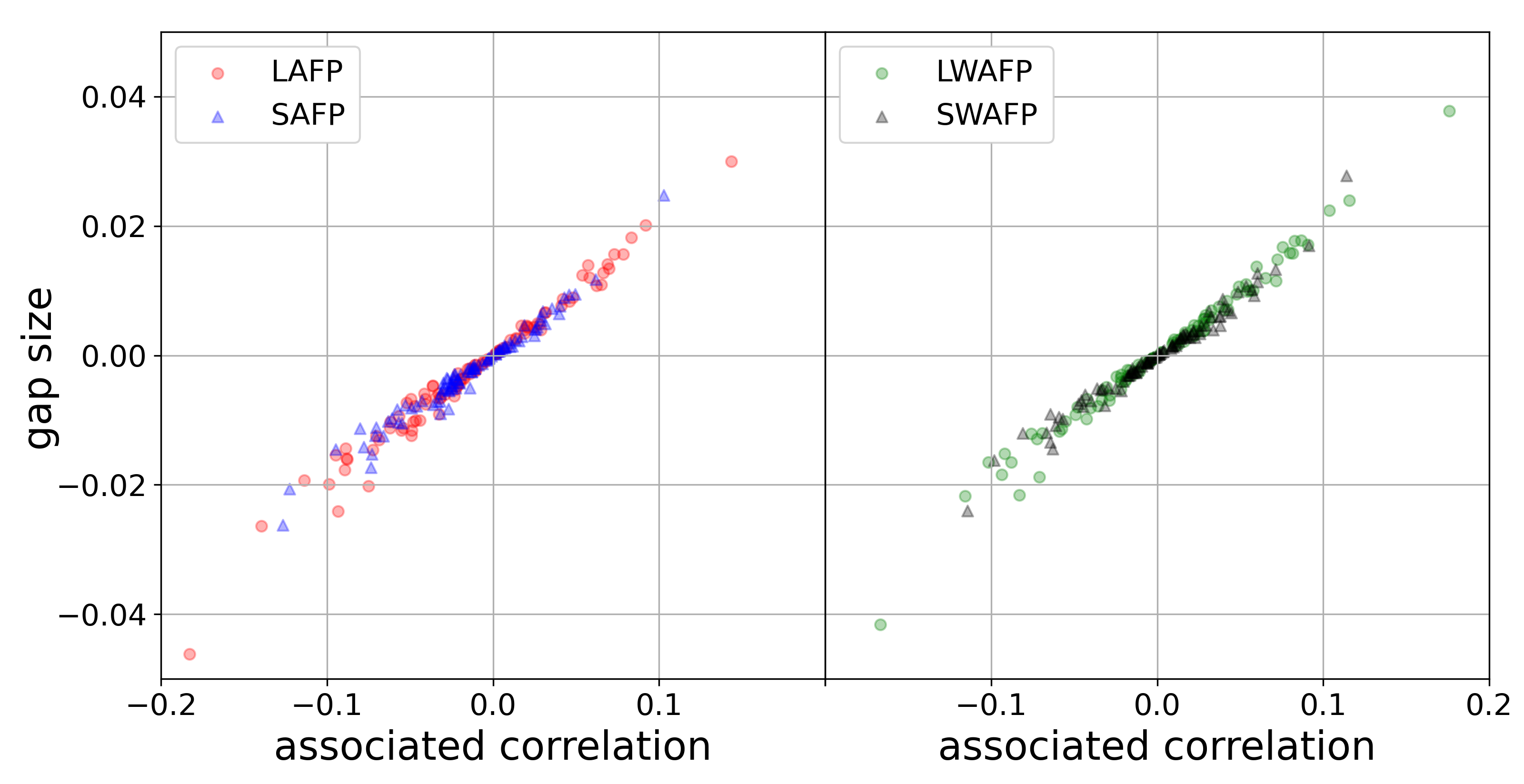}
\caption{Gap sizes and their associated correlations for the Facebook100 data. Like in Figure~\ref{fig:j0} that looked at random graphs, the LAFP gap signs follow the $r_{d,a}$ signs, the SAFP gap signs follow the $r_{\delta,a}$ signs, the LWAFP gap signs follow the $r_{w,a}$ signs, and SWAFP gap signs follow the $r_{\gamma,a}$ signs. But here the correlation between the $x$-axis quantity and the $y$-axis quantity is lower than $0.9994$ (but still very high): $0.986$ for $r_{d,a}$ and $g_{LAFP}$ (red circles), $0.983$ for $r_{\delta,a}$ and $g_{SAFP}$ (blue triangles), $0.990$ for $r_{w,a}$ and $g_{LWAFP}$ (green circles), $0.986$ for $r_{\gamma,a}$ and $g_{SWAFP}$ (black triangles).
\label{fig:fb1}}
\end{figure}

We computed $r_{d,a}$, $r_{\delta,a}$, $r_{w,a}$ and $r_{\gamma,a}$, and the LAFP, SAFP, LWAFP and SWAFP gap sizes for each network and created Figure~\ref{fig:fb1} which is analogous to Figure~\ref{fig:j0}. The $400$ LFP, SFP, LWFP and SWFP gaps are all positive as well. LAFP fails (the gap is negative) in $60$ out of $100$ cases, SAFP fails in $66$, LWAFP fails in $50$ and SWAFP fails in $49$. None of the gaps are zero. The failure rate is not quite 50\% like we would see with simulated networks and pure random attribute assignment where expected $r_{d,a}$, $r_{\delta,a}$, $r_{w,a}$ and $r_{\gamma,a}$ would all be zero. For Facebook100 and our attribute construction, average $r_{d,a}=-0.0124$, average $r_{\delta,a}=-0.0148$, average $r_{w,a}=-0.00269$ and average $r_{\gamma,a}=-0.000119$. Clearly, if these averages were higher or lower, we'd expect the associated paradox to fail less/more often.

Figure~\ref{fig:fb1} shows that the gap and its associated correlation are themselves highly correlated even in a real-world example. One should note, though, that in principle the gap-correlation correlation could be much lower than $1$ (if the networks in question come from different contexts, etc.). As an example, take the network in Figure~\ref{fig:example}. As we saw in Section~\ref{sec:example}, the LAFP gap is equal to $-\frac{1}{4}$ for this network, and the degree-attribute correlation $r_{d,a}$ is equal to $-0.866$. Now, alter that network by changing the attribute of node $A$ ($a_A$) from $2$ to $3$ and by adding node $D$ that has attribute $0$ and is attached only to $B$. (See the Supplementary Information for an image of this network.) The new LAFP gap is equal to mean$([0,3,1,0,0,0])-$mean$([3,0,1,0])=-\frac{1}{3}$ and the new $r_{d,a}$ is equal to $-0.471$. Now, the vector of gaps for these two networks, $(-\frac{1}{4},-\frac{1}{3})$, and the vector of $r_{d,a}$'s for these two networks, $(-0.866,-0.471)$, have correlation $-1$. What's important is that within any single network, the attribute-based gap signs are equal to the signs of their associated correlations as we showed in the theory part of the paper (Section~\ref{sec:efp}).

\subsubsection{Configuration model}
An important way in which a real social network is different from one generated by $G_{n,p}$ is the presence of homophily/clustering. We can find a \say{middle ground} between a random network and a Facebook100 network by removing the social structure from the Facebook100 network with a configuration model. 

A configuration model takes in a sequence of nodes and a node-indexed degree sequence and creates edges between nodes such that each node has the degree specified in the degree sequence. Using that, we can retain nodes' degrees but make their friendships more random. Note that the configuration model may generate self-loops and parallel edges. We will remove self-loops and duplicate edges, and then check that the graph is still connected. If it's not, we will redo the procedure. If it is, we will settle on this graph and create a new degree sequence from it (some degrees may be lower since we may have removed some edges).

We still want the graph to be weighted, and we want the weights to be distributed randomly too (and not based on node similarity like they previously were). For each Facebook100 network, we know the proportion of edges that were of weight $2$ (call it $p_{2}$). For each edge in the configuration model network, we assign it a weight of $2$ with probability $p_2$, and a weight of $1$ with probability $1-p_2$.

As for attributes, our original attributes---the proportion of friends that have the same gender value---are based on the social structure of the graph. Rewiring the edges will change that number for each node. Our goal with the configuration model, though, is to see how removing the social structure alone changes the gaps and the correlations, so we will keep each node's attribute unchanged from the original network to the configuration model.

Once we run the configuration model for every network, we can create plots that compare the original gap sizes/correlation values to the new ones. All such plots are shown in the Supplementary Information. We find that the \textbf{list} gaps and correlations mostly remain very similar to the originals (with the scatter plots being close to the $x=y$ line), but the \textbf{singular} gaps and correlations trend upward (the slope of a best-fit line is greater than 1). In terms of the network topology, that is mirrored in $r_{d,\delta}$ and $r_{w,\gamma}$ being much higher for the configuration model graphs than for the original graphs. Since delta and gamma describe a node's immediate neighborhood, both $r_{d,\delta}$ and $r_{w,\gamma}$ are measures of local properties in a graph. It makes sense that changing them affects the singular gaps (and their associated correlations) which deal with node-specific second-order values, and doesn't affect the list gaps (and their associated correlations) which are about network-level second-order values. Why removing social structure from a social-network graph (and why raising $n$ or $p$ for $G_{n,p}$) raises $r_{d,\delta}$ and $r_{w,\gamma}$ is beyond the scope of this paper, but we provide these results for completeness and as a precursor for potential future research.

\section{Conclusion}
\label{sec:conclusion}
In this paper we showed that even though the Friendship Paradox may initially be a confusing concept, it and its basic generalizations are very tractable mathematically, and allowing for edge weights and arbitrary node attributes can be covered in two extensions---one network-level and one node-level. Additionally, for the attribute-based versions of the paradox, we showed how the gap signs follow simple correlation-based rules. We have also provided simulation and data results that further one's understanding of the theory behind the $8$ versions of the Friendship Paradox.

\section*{Acknowledgements}
Thanks to Aleksandr Berdnikov for helpful comments.

\section*{Data availability} 

The dataset we have used, Facebook100, is introduced in a paper by Traud et al. and is available from the authors of that work: \href{http://sciencedirect.com/science/article/abs/pii/S0378437111009186}{sciencedirect.com/science/article/abs/pii/S0378437111009186}

\section*{Supplementary information}
The supplementary information is available via \textit{Scientific Reports} at\\ \underline{\href{https://www.nature.com/articles/s41598-024-63167-9}{nature.com/articles/s41598-024-63167-9}}.


\begin{thebibliography}{10}

\bibitem{feld1991your}
S.~L. Feld. Why your friends have more friends than you do. {\it American Journal of Sociology\/} {\bf 96}, 1464 (1991).

\bibitem{hodas2013friendship}
N.~O. Hodas, F.~Kooti, K.~Lerman. Friendship paradox redux: Your friends are more interesting than you. {\it Seventh International AAAI Conference on Weblogs and Social Media\/} (2013).

\bibitem{berenhaut2019friendship}
K.~S. Berenhaut, H.~Jiang. The friendship paradox for weighted and directed networks. {\it Probability in the Engineering and Informational Sciences\/} {\bf 33}, 136 (2019).

\bibitem{eom2014generalized}
Y.-H. Eom, H.-H. Jo. Generalized friendship paradox in complex networks: The case of scientific collaboration. {\it Scientific reports\/} {\bf 4}, 4063 (2014).

\bibitem{ugander2011anatomy}
J.~Ugander, B.~Karrer, L.~Backstrom, C.~Marlow. The anatomy of the facebook social graph. {\it arXiv preprint arXiv:1111.4503\/}  (2011).

\bibitem{momeni2016qualities}
N.~Momeni, M.~Rabbat. Qualities and inequalities in online social networks through the lens of the generalized friendship paradox. {\it PloS one\/} {\bf 11}, e0143633 (2016).

\bibitem{cantwell2021friendship}
G.~T. Cantwell, A.~Kirkley, M.~Newman. The friendship paradox in real and model networks. {\it Journal of Complex Networks\/} {\bf 9}, cnab011 (2021).

\bibitem{evtushenko2023node}
A.~Evtushenko, J.~Kleinberg. Node-based generalized friendship paradox fails. {\it Scientific reports\/} {\bf 13}, 2074 (2023).

\bibitem{hardy1952inequalities}
G.~H. Hardy, J.~E. Littlewood, G.~P{\'o}lya, G.~P{\'o}lya, {\it et~al.\/}. {\it Inequalities\/} (Cambridge university press, 1952).

\bibitem{kramer2016multistep}
J.~B. Kramer, J.~Cutler, A.~Radcliffe. The multistep friendship paradox. {\it The American Mathematical Monthly\/} {\bf 123}, 900 (2016).

\bibitem{granovetter1973strength}
M.~S. Granovetter. The strength of weak ties. {\it American journal of sociology\/} {\bf 78.6}, 1360 (1973).

\bibitem{traud2012social}
A.~L. Traud, P.~J. Mucha, M.~A. Porter. Social structure of facebook networks. {\it Physica A: Statistical Mechanics and its Applications\/} {\bf 391}, 4165 (2012).

\bibitem{altenburger2018monophily}
K.~M. Altenburger, J.~Ugander. Monophily in social networks introduces similarity among friends-of-friends. {\it Nature human behaviour\/} {\bf 2}, 284 (2018).

\end{thebibliography}
\end{document}